\documentclass[12pt]{revtex4}

\usepackage[]{graphicx}
\usepackage{amsmath}
\usepackage{amsfonts}
\usepackage{amssymb}
\usepackage{makeidx}
\usepackage{epstopdf}

\newcommand{\beq}{\begin{equation}} \newcommand{\eeq}{\end{equation}}
\newcommand{\bea}{\begin{eqnarray}} \newcommand{\eea}{\end{eqnarray}}
\newcommand{\bear}{\begin{eqnarray*}} \newcommand{\eear}{\end{eqnarray*}}
\newcommand{\lb}{\label} 
\newcommand{\rf}[1]{(\ref{#1})}

\begin{document}

\title {The spectral gap and the dynamical critical exponent of an exact 
solvable probabilistic cellular automaton}

\author{M.~J. Lazo$^1$, A. A. Ferreira$^2$, and \ F. C. Alcaraz$^3$}

\affiliation{$^1$Instituto de Matem\'atica, Estat\'{\i}stica e F\'{\i}sica, Universidade Federal do Rio Grande, 96.201-900 Rio Grande, Rio Grande do Sul, Brazil}
\affiliation{$^2$Laborat\'orio de F\'\i sica Te\'orica e Computa\c c\~ao Cient\'\i fica, Universidade Federal de S\~ao Paulo, Campus Diadema,  09913-030 S\~ao Paulo, S\~ao Paulo, Brazil}
\affiliation{$^3$ Instituto de F\'{i}sica de S\~ao Carlos, Universidade de S\~ao Paulo, Caixa Postal 369, 13560-970 S\~ao Carlos, S\~ao Paulo, Brazil}

%
%

\begin{abstract}

We obtained the exact solution of a probabilistic cellular automaton related to the diagonal-to-diagonal transfer matrix of the six-vertex model on a square 
lattice. The model describes the flow of ants (or particles), traveling on an 
one-dimensional lattice whose sites are small craters containing sleeping or 
awake ants (two kinds of particles). We found the Bethe ansatz equations and the spectral gap for the time-evolution operator of the cellular automaton. From the spectral gap we show that in the asymmetric case it belongs to the Kardar-Parisi-Zhang (KPZ) universality class, exhibiting  a dynamical critical exponent value $z=\frac{3}{2}$.  This result is also obtained from a direct Monte Carlo simulation, by evaluating the lattice-size dependence of the decay time to 
the stationary state.

\end{abstract}


\maketitle

\section{Introduction}

The six-vertex model was introduced in 1931 by Pauling in order to explain the residual entropy of the ice at zero temperature. The model turns out to be of great interest for physicists and mathematicians of many-body systems due to its 
exact integrability \cite{lieb1}. The row-to-row transfer matrix of the six-vertex model is the generating function for an infinite set of commuting non-trivial charges in involution \cite{tarasov}. The anisotropic Heisenberg chain, or the so called XXZ quantum chain, is one of these conserved charges. Actually, a quantum system is integrable whenever its Hamiltonian  belongs to an infinite set of commuting operators. The exact integrability of the  XXZ quantum chain is then a consequence of the infinite number of commuting charges generated by the six-vertex model. For this reason the six-vertex model is considered as a paradigm of exact integrability in statistical mechanics \cite{lieb,sutherland-yang-yang,baxter,gaudin}.

On the other hand, the representation of interacting stochastic particle dynamics in terms of quantum spin systems produced interesting and fruitful interchanges among the area of equilibrium and non-equilibrium statistical mechanics. The connection among these areas follows from the similarity between the master equation describing the time-fluctuations on non-equilibrium stochastic problems and the quantum fluctuations of equilibrium quantum spin chains~\cite{lushi,shu1,alcrit1,alcrit2,alcaraz1,stinch1,krebs1,shu2,mario1,spohn,kim,dasmar,sasawada,PRE,
der3,ligget,BJP1,BJP2,schu-domb}. The simplest example is the asymmetric diffusion of hard-core particles on the one dimensional lattice (see~\cite{der3,ligget,schu-domb} for reviews), where the time fluctuations are governed by the time evolution operator that coincides with the exact integrable XXZ quantum chain in its ferromagnetically ordered regime.

An important consequence of the above mentioned mathematical connection between quantum chains and interacting stochastic problems is that, unlike the area of non-equilibrium interacting systems where very few models are fully solvable, there exists a huge family of quantum chains appearing in equilibrium problems that are integrable by the Bethe ansatz on its several formulations (see~\cite{baxter,revkore,revessler,revschlo} for reviews). In these cases, the Bethe ansatz enable us to obtain exactly the complete spectrum of the master equation describing the time-fluctuations on the non-equilibrium stochastic problem. From the spectrum we can compute important properties of the system. As for example, the relaxation time to the stationary state which depends on the system size $L$. It 
 satisfies a scaling relation $\tau_L\sim L^{z}$, where the dynamical exponent $z$ can be obtained from  the finite-size dependence of the real part of spectral gap between of the two leading eigenvalues ruling the long time regime.

In addition, although less explored in the literature, there is also a non-trivial connection  between the transfer matrix operator of two-dimensional vertex models in thermal equilibrium and the time evolution operator in discrete time Markov chains \cite{schu-domb}. Examples of this connection are given by  the six-vertex model. It  describes a non-local parallel diffusion of hard-core particles \cite{spohn,kim}, and also  the time-evolution operator of a two steps local cellular automaton describing a half-parallel dynamics of asymmetric diffusion of hard-core particles \cite{schu-domb,KDN}. 
Another application is  given  by  the   ten-vertex model introduced in \cite{10vertex} to describe a parallel dynamics of a traffic flow stochastic cellular automaton.

The connection of the KPZ dynamics with the row-to-row transfer matrix of 
the six-vertex model
 was first observed by
Gwa and Spohn \cite{spohn}.
 This result for the six-vertex model on a cylinder was confirmed by several works including the analytical computation from Bukman and Shore \cite{bukman}.
Furthermore, more recently a rigorous proof of the KPZ exponent was also
obtained for the six-vertex model on a quadrant with parameters on the
stochastic line by Borodin, Corwin and Gorin \cite{borodin}.
It is interesting also to mention that in  \cite{halpin} we can find a 
recent review of theoretical and experimental realizations 
of the KPZ dynamics.

In the present work we study a simple one-dimensional probabilistic cellular automaton related to the diagonal-to-diagonal transfer matrix of the 
six-vertex model. This model is inspired in the idea of Alcaraz and Bariev \cite{10vertex} and describes the flow of particles, ants, automobiles, or some other conserved quantity in a one-dimensional discrete chain. Despite the diagonal-to-diagonal six-vertex model being known to be solvable for quite a long time \cite{bariev-81}, to our knowledge the spectral gap for this model with Boltzmann weights associated to stochastic processes was not considered previously. In this work we calculate the spectral gap for this model and found that the associated probabilistic cellular automaton belongs to the Kardar-Parizi-Zhang (KPZ) universality class with a dynamical exponent $z=\frac{3}{2}$ \cite{KPZ}. 

The layout of this papers is as follows. In Section II we introduce 
 the probabilist cellular automaton and show its correspondence 
with the  row-to-row transfer matrix of the six vertex model. In section III we solve exactly the model using a matrix-product ansatz. In section IV we 
calculate  
numerically the spectral gap related to the highest eigenvalues of the time-evolution operator and obtain the dynamical critical exponent. In Section V we 
perform Monte Carlo simulations of the cellular automaton. Finally in section VI we conclude our paper with a summary of the main results.


\section{The stochastic cellular automaton and its connection with the exactly integrable six-vertex mode}

The main ingredients of a cellular automaton are the associated configuration 
space and the dynamical rules defining its time evolution. The cellular 
automaton we consider is defined on a $L$-site chain where at each site  we 
attach a variable that have four possible values, and the configuration space 
has  then $4^L$ components. 
We can associate to the four possibilities of site occupation 
arbitrary objects. We chose here a simple formulation
of the automation where the four possibilities corresponds to a vacant site, 
a site occupied by a single awake or sleeping ant, and a site double 
occupied by an awake and a sleepy ant. We should stress here that the 
probabilist cellular automaton we introduce is not intended to describe the 
realistic motion of real ants. For a more realistic motion we should also 
consider the effect of the pheromones in the ants motion, like e.g., in
\cite{nishinari}. 

\begin{figure}[hbt]
\centerline{\includegraphics[width=\linewidth]{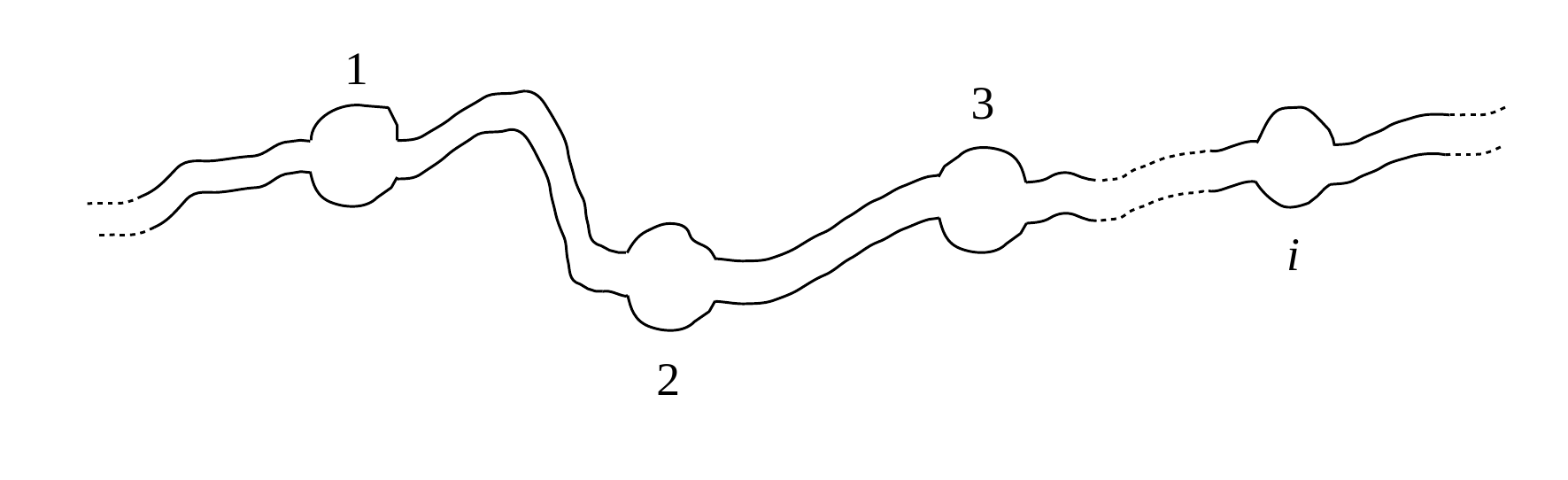}}
\caption{An illustrative example of a one dimensional set of  tunnels 
connecting  small craters.} \label{cave}
\end{figure}

We consider  $n$ ants traveling in a one dimensional tunnel that 
links $L$ small craters (sites), as shown in Fig.~\ref{cave}. 
On the craters the ants are on a sleeping or awake state. 
The craters are so small 
that at most it  supports two ants and no more  than one awake ant.
 The four possible occupations of a given crater are then: no ants, one ant (sleeping or awake) or two ants (one sleeping and another awake). 

\begin{figure}[hbt]
\centerline{\includegraphics[width=\linewidth]{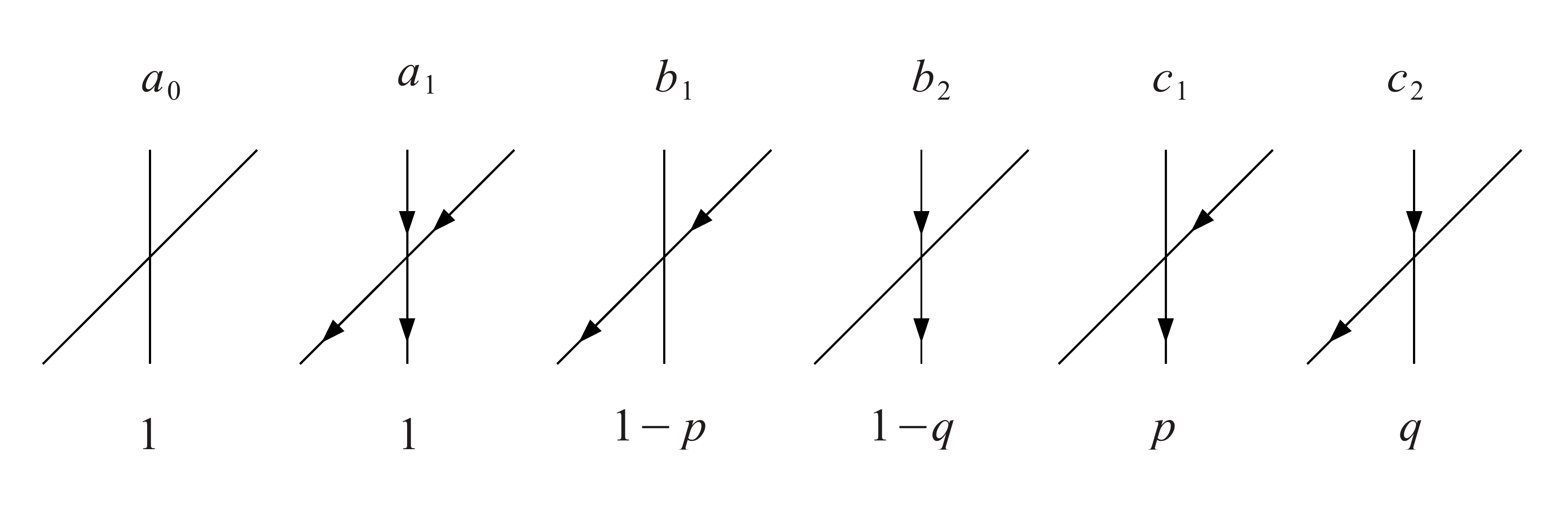}}
\caption{The fugacities of the six allowed vertex configurations of the equivalent vertex model. They correspond to the probabilities of motion in the probabilistic cellular automaton.} \label{transformacao}
\end{figure}

The dynamical rules are defined as follows. At a given unit of time all the ants travel to the next nearest crater to the left, or stay in the same crater. The travel happens with the following rules:
\begin{itemize}
\item[a)] If the ant is awake and alone in the crater it sleeps with probability $p$ or travels to the left crater with the probability $1-p$.
\item[b)] If when the ant arrives to a crater it finds another ant sleeping, since they do not want a dispute (the connecting channels among the craters only allow a single traveling ant) with probability $1$ the ant travels to the left leaving the other one sleeping.
\item[c)] If in a given crater there is a sleeping ant with a probability $q$ it awakes and travels to the left and with probability $1-q$ stay sleeping.
\end{itemize}
The allowed motions, in a unit of time, can be represented by the arrow configurations of the six 
vertices  shown in Fig. \ref{transformacao}.
The motions in Fig.~\ref{transformacao} are from  the top to the bottom  of the 
figure. 
 A vertical down arrow ($\downarrow$) represents an ant sleeping 
while the  diagonal arrow ($\swarrow$) an awake ant, and lines with no arrows 
indicate the absence of  ants.  For simplicity, we consider periodic boundary condition. The conservation of the number of ants is then a consequence of  the ice type rules  defining the vertices, that demand an equal number of outgoing and incoming arrows in a given site. 

\begin{figure}[hbt]
\centerline{\includegraphics[width=\linewidth]{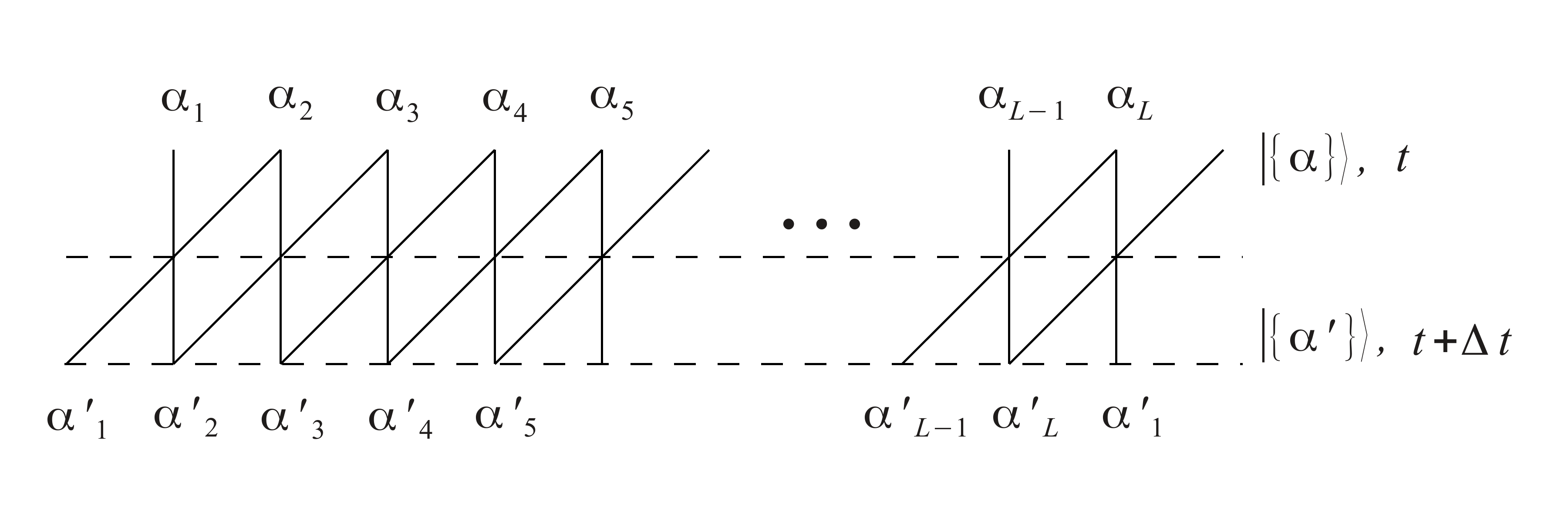}}
\caption{Pictoric representation of the diagonal-to-diagonal transfer matrix. The variables $\alpha_j$ and $\alpha'_j$ specify the possible arrow configurations at site position $j$ at times $t$ and $t+\Delta t$, respectively.
} \label{matrix}
\end{figure}

Let us denote as $\alpha_j$ the possible occupations of the crater $j$ 
 by $\alpha_j=0,1,2$ and 3, that  indicates that there is no ant, a sleeping ant, an awake ant, and a sleeping and awake and, respectively.
 The probability distribution  $P(\{\alpha\},t)$  of finding the system in the state $|\{\alpha\}\rangle=|\alpha_1, \alpha_2,...,\alpha_j,..., \alpha_L\rangle$ at time $t$ is given by the components of the probability vector.
\beq
\lb{a1}
|P(\{\alpha\},t)\rangle=\sum_{\{\alpha\}}P(\{\alpha\},t)|\{\alpha\}\rangle,
\eeq
where the summation is performed over all possible configurations $\{\alpha\}=\{\alpha_1,...,\alpha_L\}$ satisfying the ice type rules. The time evolution 
of  the stochastic cellular automaton is given by the master equation:
\beq
\lb{a2}
|P(\{\alpha\},t+\Delta t)= T_{D-D} |P(\{\alpha\},t),
\eeq
where the transition matrix $T_{D-D}$ corresponds to the 
 diagonal-to-diagonal transfer matrix 
of a six-vertex model with fugacities $a_0,a_1,b_1,b_2,c_1,c_2$ related, 
respectively,  to the 
probabilities $1,1,1-p,1-q,p,q$ of the cellular automaton (see Fig.~\ref{transformacao}). The components 
$\langle\{\alpha\}|T_{D-D}|\{\alpha'\}\rangle$
 give the probability of motion from configuration $\{\alpha\}$ to 
$\{\alpha'\}$ and are given by the product of the vertex fugacities connecting 
the configurations:
\beq
\lb{a3}
\langle\{\alpha\}|T_{D-D}|\{\alpha'\}\rangle=a_0^{n_0}a_{1}^{n_1}b_1^{n_2}b_2^{n_3}c_1^{n_4}c_2^{n_5},
\eeq
where $n_0,n_1,n_2,n_3,n_4,n_5$ are the numbers of vertices with fugacities $a_0,a_1,b_1,b_2,c_1,c_2$, respectively (see Fig. \ref{matrix}).

The spectral properties of the  six-vertex model are characterized by the parameter \cite{baxter}:
\begin{eqnarray}
\Delta=\frac{b_1b_2-c_1c_2+a_1a_0}{2a_1b_2},
\end{eqnarray}
that for the cellular automaton is given by 

\begin{eqnarray}
\Delta =\frac{1-\frac{(p+q)}{2}}{1-q}.
\end{eqnarray}


\section{Diagonalization of the transfer matrix}

In this section we solve the eigenvalue equation for the diagonal-to-diagonal transfer matrix $T_{D-D}$ introduced in the previous section. As a consequence of the conservation of arrows, and due to the periodic boundary condition, the transfer matrix $T_{D-D}$ can be split into block disjoint sectors labeled by the number $n$ $(n=0,1,\ldots,2L)$ of arrows and by the momentum $P=\frac{2\pi}{L}j$ $(j=0,1,\ldots,L).$ We want to solve, in each of these sectors, the eigenvalue equation
\begin{eqnarray}
\label{eq:auto}
\Lambda_{n,P}|\Psi_{n,P}>=T_{D-D}|\Psi_{n,P}>,
\end{eqnarray}
where $\Lambda_{n,P}$ and $|\Psi_{n,P}>$ are the eigenvalues and eigenvectors of $T_{D-D}$, respectively. These eigenvectors can be written in general as
\begin{eqnarray}
\label{eq:auto2}
|\Psi_{n,P}>=\sum_{\{x\}}\sum_{\{\alpha\}}^{(*)}\phi^{P}_{\alpha_1,\ldots,\alpha_n}(x_ 1,\ldots,x_n)|x_1,\alpha_1;\ldots;x_n,\alpha_n>,
\end{eqnarray}
where $\phi^{P}_{\alpha_1,\ldots,\alpha_n}(x_1,\ldots,x_n)$ is the amplitude corresponding to the arrows configuration where $n$ arrows of type $\alpha_1,\ldots,\alpha_n$ are located at positions $x_1,\ldots,x_n$, respectively (where $\alpha_j=1$ correspond to a vertical arrow and $\alpha_j=2$ to a diagonal arrow). Finally, the symbol $(*)$ in the last equation means that the sums in $\{x\}$ and $\{\alpha\}$ are restricted to the sets obeying the ice rule.

Since $|\Psi_{n,P}>$ is also an eigenvector of the translation operator with momentum $P=\frac{2\pi}{L}j$ ($j=0,...,L-1$), the amplitudes satisfy
\begin{eqnarray}
\label{eq:auto3}
\frac{\phi^{P}_{\alpha_1,\ldots,\alpha_n}(x_ 1,\ldots,x_n)}{\phi^{P}_{\alpha_1,\ldots,\alpha_n}(x_ 1+1,\ldots,x_n+1)}=e^{-iP},
\end{eqnarray}
for $x_n\leq L-1,$ whilst for $x_n=L$
\begin{eqnarray}
\label{eq:auto4}
\frac{\phi^{P}_{\alpha_1,\ldots,\alpha_n}(x_ 1,\ldots,L)}{\phi^{P}_{\alpha_1,\ldots,\alpha_n}(1,x_ 1+1\ldots,x_{n-1}+1)}=e^{-iP}.
\end{eqnarray}

The solution of the eigenvalue equation \rf{eq:auto} can be obtained by an appropriate ansatz for the unknown amplitudes $\phi^{P}_{\alpha_1,\ldots,\alpha_n}(x_ 1,\ldots,x_n)$. Although the model can be solved by the coordinate Bethe ansatz \cite{bethe}, we are going to formulate a new matrix product ansatz (MPA) \cite{alclazo1,alclazo2,alclazo3} due its simplicity and unifying implementation for arbitrary systems. This new MPA introduced in \cite{alclazo1,alclazo2,alclazo3} can be seen as a matrix product formulation of the coordinate Bethe ansatz and it is suited to describe all eigenvectors of integrable models, including spin chains \cite{alclazo1,alclazo2,alclazo3}, stochastic models \cite{alclazo4,LazoAnd1,LazoAnd2} and transfer matrices \cite{anderson-pre,lazo,alclazo5}. According to this ansatz, there is a correspondence between the amplitudes of the eigenvectors and matrix products among matrices obeying  special algebraic relations. In the present work, in order to formulate the MPA we make a one-to-one correspondence between the configurations of arrows and the products of matrices. The matrix product associated to a given arrow configuration is obtained by associating a matrix $E$ to the sites with no arrow, a matrix $A^{(\alpha)}$ to the sites with a single arrow of type $\alpha$ $(\alpha=1,2)$, and finally the matrix $A^{(1)}E^{-1}A^{(2)}$ for sites with two arrows. The unknown amplitudes in \rf{eq:auto2} are obtained by associating them to the matrix product ansatz
\beq
\label{eq:pala}
\phi^{P}_{\alpha_1,\ldots,\alpha_n}(x_ 1,\ldots,x_n) \Longleftrightarrow E^{x_1-1}A^{(\alpha_1)}E^{x_2-x_1-1}A^{(\alpha_2)}\cdots E^{x_n-x_{n-1}-1}A^{(\alpha_n)}E^{L-x_n},
\eeq
if there are no sites with two arrows, and by associating
\beq
\label{eq:pala2}
\begin{split}
&\phi^{P}_{\alpha_1,\ldots,\alpha_n}(x_ 1,\ldots,x_n) \Longleftrightarrow E^{x_1-1}A^{(\alpha_1)}E^{x_2-x_1-1}A^{(\alpha_2)}\cdots\\
&\;\;\;\;\;\;\;\;\;\;\;\;\;\;\;\;\;\;\;\; \cdots E^{x_j-x_{j-1}-1}A^{(1)}E^{-1}A^{(2)}E^{x_{j+1}-x_j-1} \cdots E^{x_n-x_{n-1}-1}A^{(\alpha_n)}E^{L-x_n}.
\end{split}
\eeq
if there are two arrows at position $x_j$. The other cases follows straightforwardly.

Actually $E$ and $A^{(\alpha)}$ are in general abstract operators with an associative product. A well defined eigenvector is obtained, apart from a normalization factor, if all the amplitudes are related uniquely, due to the algebraic relations (to be fixed) among the matrices $E$ and $A^{(\alpha)}$. Equivalently, the correspondences \rf{eq:pala} and \rf{eq:pala2} implies that, in the subset of words (products of matrices $A^{(\alpha)}$ and $E$) there exists only a single independent word ("normalization constant"). The relation between any two words is a $c$ number that gives the ratio between the corresponding amplitudes. Moreover, since the eigenvectors have a well defined momentum $P=\frac{2\pi}{L}l$ $(l=0,\ldots,L-1)$, the relations \rf{eq:auto3} and \rf{eq:auto4} imply the following constraints for the matrix products appearing in the ansatz \rf{eq:pala}
\beq
\label{mom1}
\begin{split}
E^{x_1-1}A^{(\alpha_1)}E^{x_2-x_1-1}A^{(\alpha_2)}&\cdots E^{x_n-x_{n-1}-1}A^{(\alpha_n)}E^{L-x_n}=\\
&e^{-iP}E^{x_1}A^{(\alpha_1)}E^{x_2-x_1-1}\cdots A^{(\alpha_n)}E^{L-x_n-1},
\end{split}
\eeq
for $x_n\leq L-1,$ and for $x_n=L$
\beq
\label{mom2}
\begin{split}
E^{x_1-1}A^{(\alpha_1)}E^{x_2-x_1-1}A^{(\alpha_2)}&\cdots E^{x_n-x_{n-1}-1}A^{(\alpha_n)}E^{L-x_n}=\\
&e^{-iP}A^{(\alpha_n)}E^{x_1-1}A^{(\alpha_1)}\cdots A^{(\alpha_{n-1})}E^{L-x_{n-1}-1}.
\end{split}
\eeq
The relations \rf{mom1} can be easily solved by identifying  the matrices $A^{(\alpha)}$ as being composed by $n$ spectral dependent matrices $A_{k_j}$ $j=(1,\ldots,n)$,
\begin{eqnarray}
\label{spmatrix}
A^{\alpha}=\sum_{j=1}^n\phi_{\alpha}^jA_{k_j}E,\;\;\;\;\alpha=1,2,
\end{eqnarray}
satisfying the algebraic relation
\begin{eqnarray}
\lb{EA}
EA_{k_j}=e^{ik_j}A_{k_j}E.
\end{eqnarray}
 By inserting \rf{spmatrix} and using \rf{EA} into \rf{mom1} we 
verify that  the spectral parameters $k_j$ ($j=1,...,n$) are related to the momentum $P=\frac{2\pi}{L}l$ $(l=0,\ldots,L-1)$ of the eigenvector:
\begin{eqnarray}
P=\sum_{j=1}^nk_j.
\end{eqnarray}
On the other hand, by inserting \rf{spmatrix} and using \rf{EA} into the boundary equations \rf{mom1} we obtain the algebraic relations
\beq
\lb{boundary}
E^{x_1}A_{k_1}E^{x_2-x_1}A_{k_2}\cdots E^{L-x_{n-1}}A_{k_n}=A_{k_n}E^{x_1-1}A_{k_1}\cdots A_{k_{n-1}}E^{L-x_{n-1}}.
\eeq

The eigenvalue equation \rf{eq:auto} give us two kinds of relations for the amplitudes \rf{eq:pala} and \rf{eq:pala2}. The first one is related to those amplitudes without multiple occupancy, and the second one is related to those amplitudes with multiple occupancy. The first kind of relations, after some algebraic manipulations following \cite{anderson-pre} (with $t=0$ in \cite{anderson-pre}, where $t$ represent an interaction parameter of the interacting five vertex model), give us the eigenvalues $\Lambda_{n,P}(k_1,\ldots,k_n)$ of the transfer matrix
\beq
\lb{eigenvalue}
\Lambda_{n,P}(k_1,\ldots,k_n)=\prod_{j=1}^n\Lambda_1(k_j),
\eeq
where
\beq
\lb{eigenvalue2}
\begin{split}
\Lambda_1(k)&=\Lambda_1^{(\pm)}(k)=\frac{a_0^{L-1}}{2}\left(b_2+b_1e^{ik}\pm\sqrt{(b_2+b_1e^{ik})^2-4e^{ik}(b_2b_1-c_2c_1)}\right)\\
&=\frac{1}{2}\left(1-q+(1-p)e^{ik}\pm\sqrt{[1-q+(1-p)e^{ik}]^2-4e^{ik}(1-p-q)}\right).
\end{split}
\eeq

On the other hand, the relations coming from the configurations where two arrows are at same position fix the commutations relations among the matrices $A_{k_j}$:
\beq
\lb{AA}
A_{k_j}A_{k_l}=s(k_j,k_l)A_{k_l}A_{k_j},
\eeq
where
\beq
\lb{S}
\begin{split}
s(k_j,k_l)&=-\frac{\Lambda_1(k_l)\Lambda_1(k_j)b_1-\Lambda_1(k_j)(b_2b_1-c_2c_1+a_1)+a_1b_2}{\Lambda_1(k_l)\Lambda_1(k_j)b_1-\Lambda_1(k_l)(b_2b_1-c_2c_1+a_1)+a_1b_2}\\
&=-\frac{\Lambda_1(k_l)\Lambda_1(k_j)(1-p)-\Lambda_1(k_j)(2-p-q)+1-q}{\Lambda_1(k_l)\Lambda_1(k_j)(1-p)-\Lambda_1(k_l)(2-p-q)+1-q}.
\end{split}
\eeq
The relations with more than two arrows at same positions are automatically satisfied, due to the associativity of the algebra of the matrices $A_{k_j}$ $(j=1,\ldots,n)$.

Finally, the up to now free spectral parameters $\{k_i\}$ are fixed by the nonlinear Bethe equations of the diagonal-to-diagonal six-vertex model \cite{bariev-81}
\beq
\lb{Bethe}
e^{ik_jL}=-\prod_{l=1}^ns(k_j,k_l),\;\;\;\;j=1,\ldots,n,
\eeq
that is obtained by imposing the consistence between boundary relations \rf{eq:auto4} and the commutation relations \rf{EA} and \rf{S} \cite{alclazo1,alclazo2,alclazo3}.

\section{The spectral gap}

In order to complete the solution of any integrable model we need to find the roots of the associated spectral parameter equations (Eq.~\rf{Bethe} in our case). The solution of those equations is in general a quite difficult problem for finite $L$. However, numerical analysis on small lattices allows us to conjecture, for each problem, the particular distributions of roots
 that correspond to the most important eigenvalues in the bulk limit $L\rightarrow \infty$. Those are the eigenvalues with larger real part in the case of transfer matrix calculations. The equations we obtained in the last section, 
up to our knowledge,  were never analyzed previously  for either finite or infinite values of $L$ in the asymmetric case $b_1\neq b_2$ and $c_1\neq c_2$ ($p\neq q$). Actually, while a detailed  numerical analysis of the Bethe ansatz equations for the row-to-row transfer matrix of the six-vertex model, or the XXZ chain, was previously done \cite{AlcBB,BBP}, the diagonal-to-diagonal model was previously analyzed only in the symmetric case and with $\Delta>1$ \cite{anderson-pre}. 
 
\begin{figure}[hbt]
\centerline{\includegraphics[width=10cm,height=8cm]{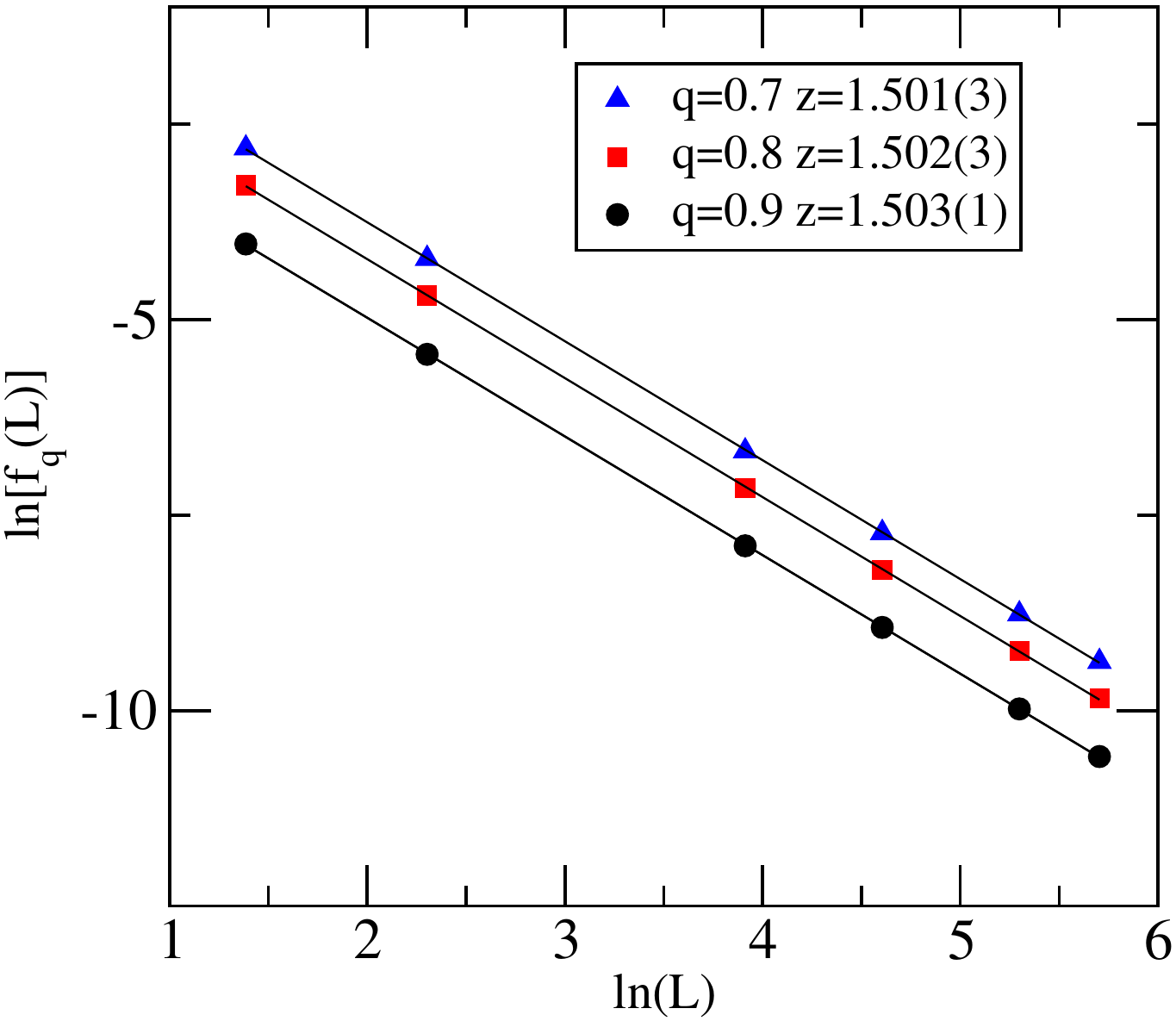}}
\caption{ The logarithm of $f_q(L)$, given in \rf{fq} as a function of $\ln L$ for the stochastic model with parameters $q=0.7,0.8,0.9$ in the in the half-filling sector $n=L$. The
dynamical exponent is close the expected value $z=\frac{3}{2}$.
} \label{gap}
\end{figure}

The corresponding Bethe equations for the diagonal-to-diagonal transfer matrix of the six-vertex model are quite different from those of the row-to-row transfer matrix. In order to simplify our analysis we are going to restrict  ourselves hereafter to the case where $p=1$ ($c_1=1$ and $b_1=1-p=0$) in a half-filled chain. In this case we have $\Delta=\frac{1}{2}$ and, from \rf{eigenvalue2} and \rf{S},
\beq
\lb{ek}
e^{ik_j}=\frac{\Lambda_1(k_j)[\Lambda_1(k_j)+q-1]}{q},
\eeq
and
\beq
\lb{S2}
S(k_j,k_l)=-\frac{[\Lambda_1(k_j)-1]}{[\Lambda_1(k_l)-1]}.
\eeq
By inserting \rf{ek} and \rf{S2} in \rf{S}, the Bethe equations can be rewritten as
\begin{eqnarray}
\label{eq:sep1}
\bigg{[}\frac{\Lambda_1(k_j)-1}{\Lambda_1(k_j)(\Lambda_1(k_j)+q-1)}\bigg{]}^L=\frac{(-1)^{L+1}}{q^L}\prod_{l=1}^L[\Lambda_1(k_l)-1]=Y,
\end{eqnarray}
where we set $n=L$ (half-filled chain). If we parametrize $Y=-a^Le^{i\theta L}$, with $a\geq 0$ and $\theta \in \left(-\frac{\pi}{L},\frac{\pi}{L}\right)$, the roots $\Lambda_1(k_j)$ of \rf{eq:sep1} are given by
\beq
\label{eq:para}
\begin{split}
&\Lambda_1(k_j)=\Lambda_1^{(\pm)}(k_j)=\frac{1-y_j(q-1)\pm\sqrt{(y_j(q-1)-1)^2-4y_j}}{2y_j},\\
&y_j=ae^{i\theta}e^{i2\pi(j-\frac{1}{2})/L}=\frac{\Lambda_1(k_j)-1}{\Lambda_1(k_j)(\Lambda_1(k_j)+q-1)},\;\;\;\; j=1,\ldots,L.
\end{split}
\eeq
For a given choice $\{\Lambda_1(k_j)\}$ of the above set of roots we have from \rf{eq:sep1} and \rf{eq:para}
\begin{eqnarray}
a^Le^{i\theta L}=\frac{(-1)^L}{q^L}\prod_{l=1}^L[\Lambda_1(k_l)-1].
\end{eqnarray}
We have solved numerically the above equations for several values of $q$. The stationary states $\Lambda^0_L\equiv\Lambda_{L,P}=1$ eigenvalue is obtained by setting $P=\sum_{j=1}^Lk_j=0$ and by choosing 
\begin{eqnarray}
\Lambda^0_L=\Lambda_1^{(+)}(k_1)\Lambda_1^{(+)}(k_2)\cdots\Lambda_1^{(+)}(k_{L-1})\Lambda_1^{(+)}(k_L),
\end{eqnarray}
where $\Lambda_1^{(+)}(k_j)$ are given by \rf{eq:para}. On the other hand, the eigenvalue having   the second largest real part belongs to the 
eigensector with momentum   $P=\sum k_j=\frac{2\pi}{L}$, and is obtained  by choosing
\begin{eqnarray}
\lb{eig1}
\Lambda^1_L\equiv\Lambda_{L,P}=\Lambda_1^{+}(k_1)\Lambda_1^{+}(k_2)\cdots\Lambda_1^{+}(k_{L-1})\Lambda_1^{-}(k_{L+1}).
\end{eqnarray}
In Table \ref{table} we display our numerical solutions for $|\Lambda^1_L|$, with $p=1$ and $q=0.7,0.8,0.9$, for several values of $L$.
\begin{table}
 \centering
\begin{tabular}{|c|c|c|c|}
 \hline
$L$ &\multicolumn{3}{|c|}{$|\Lambda^1_L(L,p=1)|$}\\
\cline{2-4} &$q=0.7$     & $q=0.8$   & $q=0.9$   \\
\hline 4    & 0.94140144138869053   & 0.96301968197766430   & 0.98238800386894887   \\
\hline 10    & 0.98538385114230209  & 0.99085409804497404   & 0.99568145771034344   \\
\hline 50    & 0.99874631037139583  & 0.99921749651303016  & 0.99962661788291240   \\
\hline 100    & 0.99955908669853510   & 0.99972484139884665    & 0.99986844854358292    \\
\hline 200    & 0.99984424258354831  & 0.99990280245468144   & 0.99995359703973385  \\
\hline 300    & 0.99991524776815632  & 0.99994711263234282   & 0.99997476777514405   \\
\hline
\end{tabular} \caption{The real part of the second largest eigenvalue of $T_{D-D}$ for lattice sizes $L=4,10,50,100,200,300$ and with parameter $p=1$ and  $q=0.7,0.8,0.9$.
}\label{table}
\end{table}
The eigenvalue with second largest  real part  \rf{eig1} determines the relaxation time and the dynamical exponent $z$. By rewritten $\Lambda^1_L= |\Lambda^1_L|e^{i\gamma}=e^{-f_{q}(L)+i\gamma(L)}$, we have
\begin{eqnarray}\label{fq}
f_{q}(L)\sim L^{-z}.
\end{eqnarray}
In Fig. \ref{gap} we show $\ln(f_q(L))$ versus $\ln(L)$ for $p=1$ and $q=0.7,0.8,0.9$. As we can see from this figure the energy gap give us  a dynamical 
critical exponent $z=\frac{3}{2}$. Consequently, the probabilistic cellular automaton introduced in Section II and the  related  asymmetric  six-vertex model belongs to the KPZ universality class \cite{KPZ}. 

\begin{figure}[hbt]
\centerline{\includegraphics[width=10cm,height=8cm]{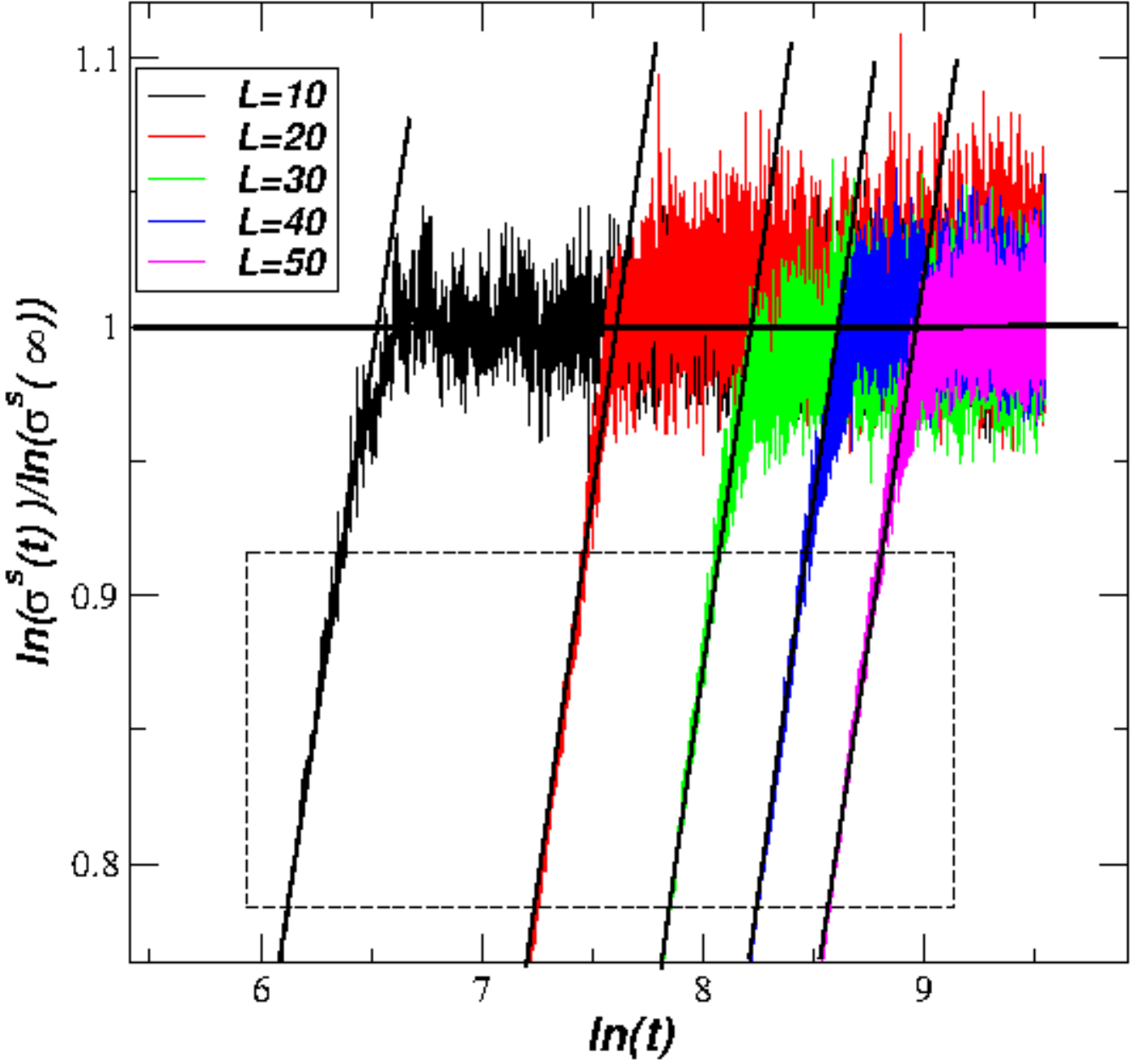}}
 \caption{The ratio $\sigma^s(t)/\sigma^s(\infty)$, as defined in \rf{variance},  of the standard deviation of 
the local density of sleeping ants, as a function of time. The results were  obtained from Monte Carlo simulations for the cellular automaton in the sector 
with $n=L$ (half-filled) ants and with parameters 
$p=1$ and $q=0.75$. At the initial state  all the $L$ craters are 
occupied by a single sleeping ant. 
The dashed box gives the region used for the linear fit to extract the 
typical decay time $\tau_L$ (see text).The results are the average of 
 $10^7$ samples  
 for each lattice size.}  
 \label{mc1}
\end{figure}
\begin{figure}[hbt]
\centerline{\includegraphics[width=10cm,height=8cm]{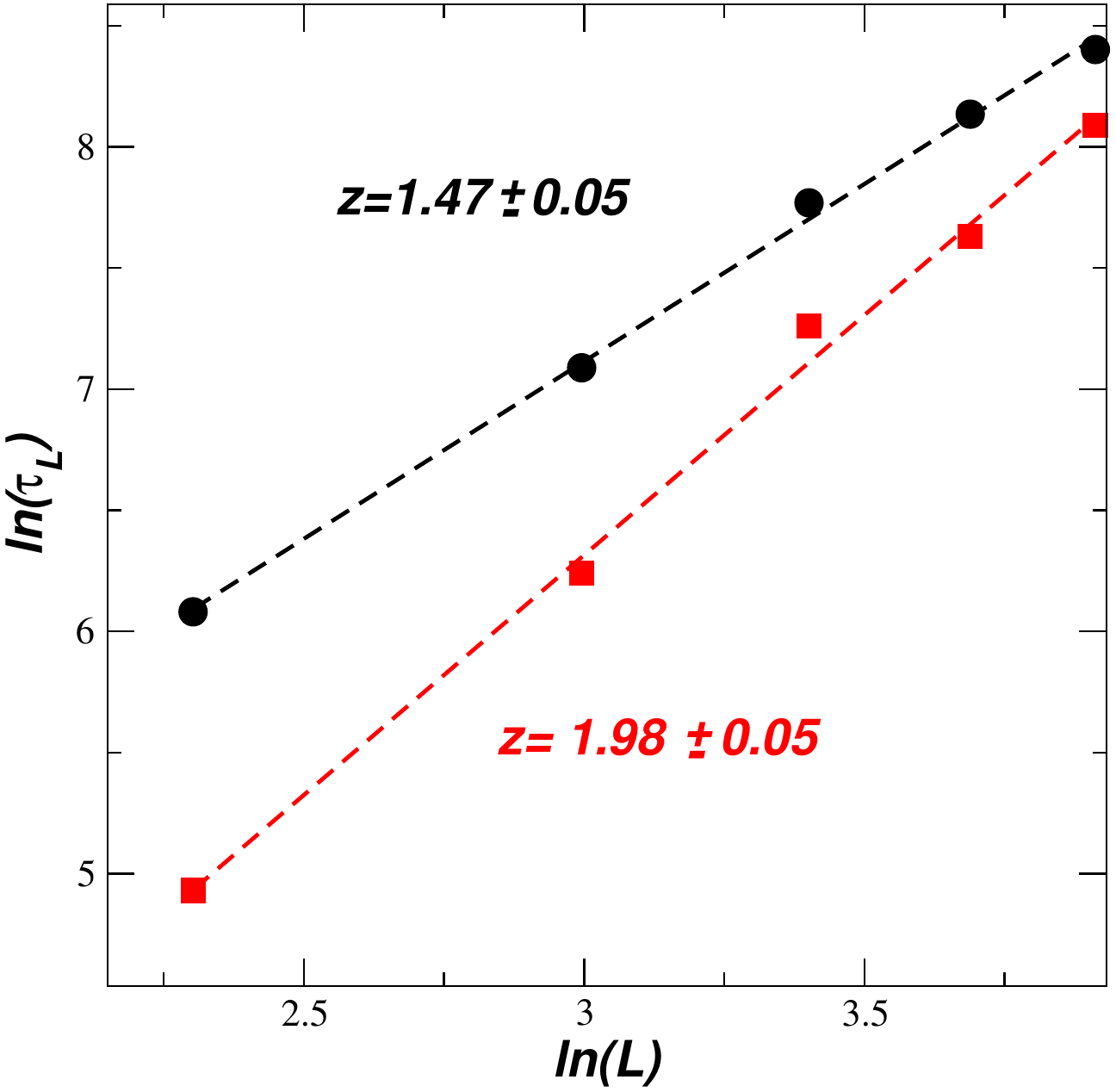}}
 \caption{The typical decay time $\tau_L$, for lattice sizes $L=20,20,30,40$ and $50$. 
The dashed box gives the region used for the linear fit to extract the 
typical decay time $\tau_L$ (see text).  The squares are obtained from 
 Fig.~{\ref{mc1}}, where $p=1$ and $q=0.75$. The circles are the 
results obtained for the symmetric case $p=q=1/2$. Both figures were obtained 
by the same procedure and number of samples. The estimated values for the dynamical critical exponent $z$ are also close to the values $3/2$ and $2$.}
 \label{mc2}
\end{figure}
\begin{figure}[hbt]
\centerline{\includegraphics[width=10cm,height=8cm]{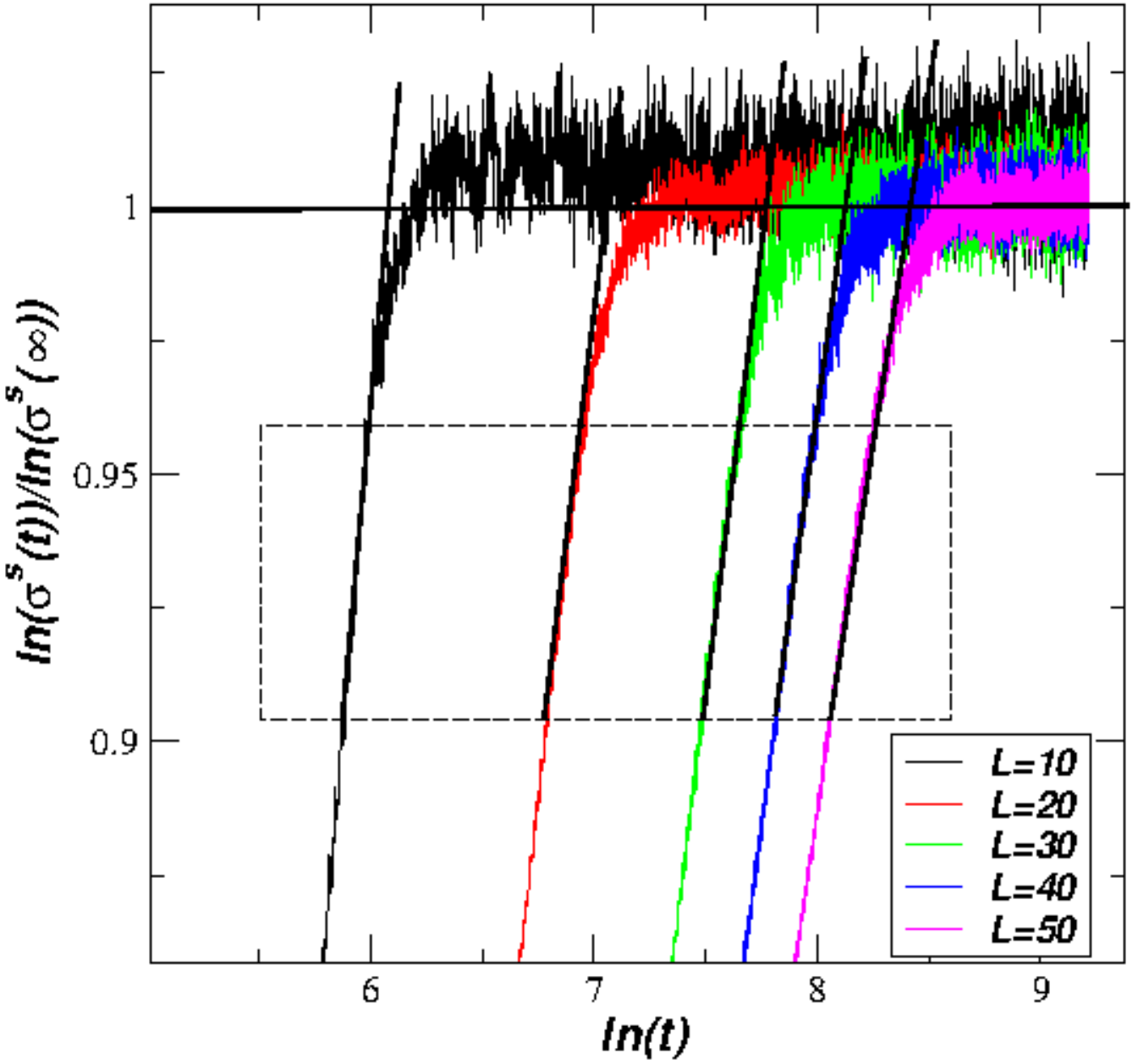}}
 \caption{The ratio $\sigma^s(t)/\sigma^s(\infty)$, as defined in \rf{variance},  of the standard deviation of 
the local density of sleeping ants, as a function of time. The results were  obtained from Monte Carlo simulations of the cellular automaton with  a small local change of the dynamical rules defined in Sec. II (see the text).  
 As in Fig.~\ref{mc1} the  parameters  are
$p=1$ and $q=0.75$ and $10^7$ samples were considered for each lattice size.
The dashed box delineate the region where a linear fit is done to extract the
typical decay time $\tau_L$ (see text).}
 \label{mc3}
\end{figure}
\begin{figure}[hbt]
\centerline{\includegraphics[width=10cm,height=8cm]{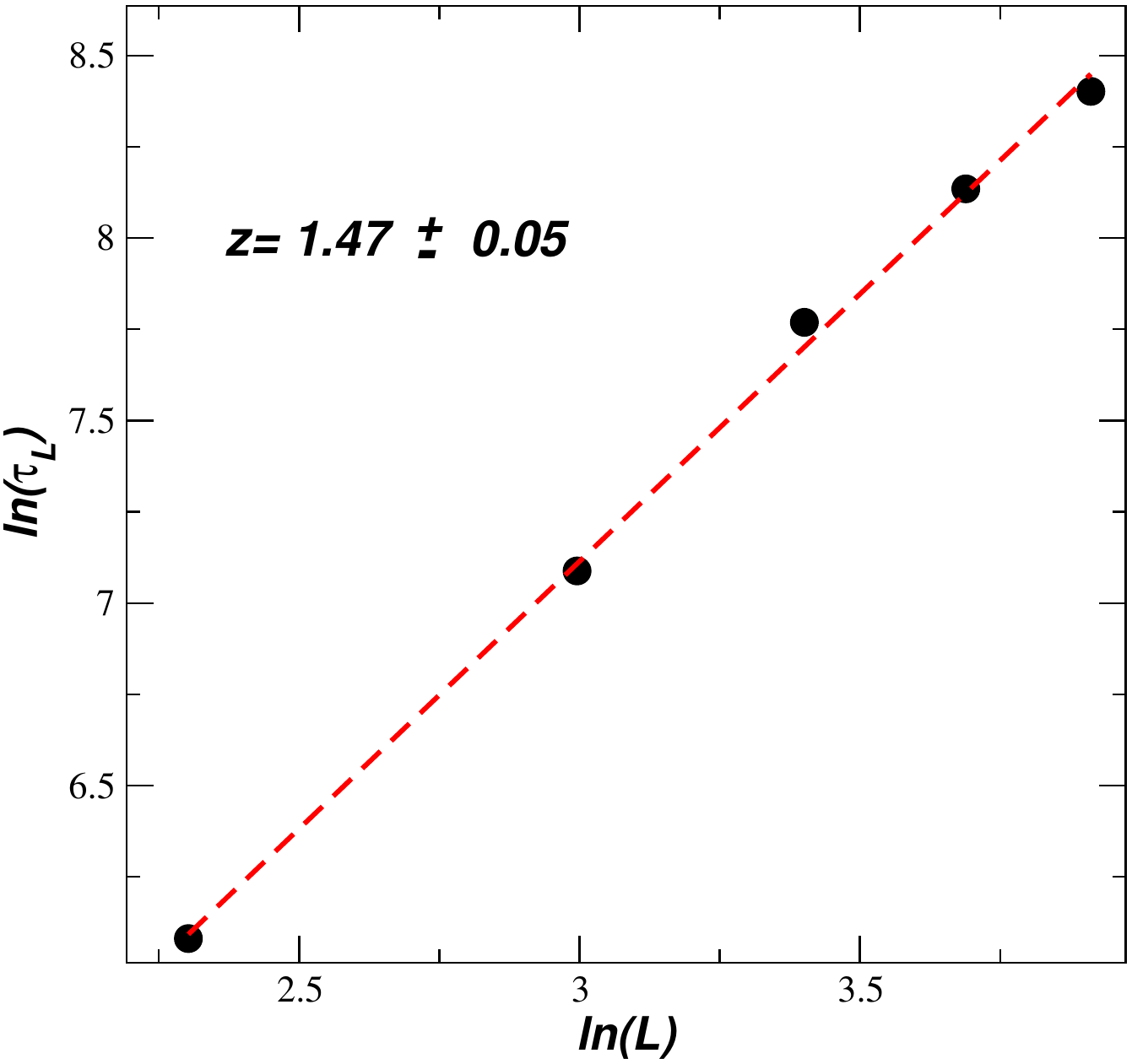}}
 \caption{The typical decay time $\tau_L$, for lattice sizes $L=20,20,30,40$ and $50$  obtained from Fig.~{\ref{mc3}}. The estimated value for the dynamical critical exponent $z$ is shown in the figure.}
 \label{mc4}
\end{figure}

\section{Monte Carlo simulations} In order to check and illustrate the 
results of last section we present some  Monte Carlo simulations (MCS) of the 
stochastic cellular automaton whose dynamic rules were defined in Sec.~II. 
 We consider, as in  last section, the case where  the number of 
ants is equal to the lattice size $n=L$ (half-filled lattice). 

It is not simple to calculate the dynamical critical exponent $z$ of models 
in the KPZ universality class by measuring  directly the decay-time of 
observables starting on a given initial condition. A known way is by 
considering the time-correlations of tagged particles (ants in the 
present case)\cite{tag1,tag2}. We verified that, alternatively,  this exponent can also be 
calculated by 
measuring 
the time evolution of the 
variance of a local operator in a given site. In particular by defining 
$n^s(i,t)$  as the occupation number (0 or 1) of a sleeping ant at site $i$ 
and time $t$, we consider the measure
\beq \label{variance}
\sigma^s(t) = \sqrt{\frac{1}{L} \sum_{i=1}^L<[n^s(i,t) - <n^s>_{\infty}]^2>},
\eeq
where $<n^s>_{\infty}$ is the asymptotic  average number of  sleeping ants 
in a  given site.

 We consider two kinds of initial states. Initially we consider a state not 
 translational invariant. We take 
 $L/2$ consecutive craters double occupied (one sleepy and one awake ant) and the remaining craters  empty. In Fig.~\ref{mc1} 
we show the ratio $\sigma^s(t)/\sigma^s(\infty)$ for lattice sizes 
$L=10,20,30,40$ and $50$, and parameters $p=1$, $q=0.75$.  We estimate from 
this figure the typical times $\tau_L$ where the systems of size $L$ 
reach the 
stationary state. These typical times are estimated from the crossing with the 
value 1 of the straight line obtained from the fitting in the region represented 
by a dash box in Fig.~\ref{mc1}.  As we can see the time 
decay $\tau_L$ increases with the lattice size. The dynamical critical 
exponent $z$ is calculated from the finite-size behavior 
$\tau_L\sim L^z$. In Fig.~\ref{mc2}  we fit the values of $\ln(\tau_L)$ with $\ln(L)$ and obtain the estimate $z =1.54 \pm 0.05$, in agreement 
with the expected value $z=3/2$.
In the case where the cellular automaton has parameters $p=q=1/2$, i. e. the 
symmetric case, we expect that the dynamics is pure diffusive.
 For the sake of comparison we also 
show in Fig.~\ref{mc2} the lattice dependence of the typical decay time 
obtained for this symmetric case. We obtain in this case 
$z=2.02 \pm 0.05$, in agreement with a standard diffusive behavior where $z=2$. 

We also consider initial states that are translational invariant, as for 
example the one where we have a single awake ant on each crater ($n=L$). 
We verified that by using the stochastic rules that define 
 the cellular 
automaton  introduced in section II it is difficult to extract the 
dynamical critical exponent $z$. In this case  the typical times where the 
system achieve the stationary state is quite small. We can however extract 
some reliable estimate of $z$ by
 introducing  a small local change in the dynamical rules, that   keeps the 
model on the same universality class of critical behavior. The modified 
dynamics keeps the number of ants conserved but
 the 
last crater ($L$) is special. If we have an awake ant at this crater 
or there is no ants coming to occupy this crater (coming from crater 1) the 
rules are the ones that define the cellular automaton (see Fig.~\ref{transformacao}). However if one ant is coming (from crater 1) 
and there is no awake 
ant at crater $L$, there is two possibilities with distinct transition probabilities. If there is 
no ant at crater $L$ with probability $p$ the arriving ant stays sleeping 
 at the crater $L$, and with probability $1-p$ goes to the 
crater $L-1$, staying  awake. If there is already a sleeping ant 
with probability one  the  awake ant that comes from crater 1, 
goes directly 
to crater $L-1$ staying awake, and  leaving at crater $L$ the  sleeping ant. 
 In Fig.~\ref{mc3} we show the ratio 
$\sigma^s(t)/\sigma^s(\infty)$ for the same lattice sizes and parameters as 
in Fig.~\ref{mc1}. In Fig.~\ref{mc4} we show the dependence of the typical 
times $\tau_L$ to reach the stationary state as a function of the lattice 
size $L$. We obtain from this last curve the estimate 
$z= 1.47 \pm 0.05$, again in agreement with the KPZ expected value $z=3/2$, 
as predicted by  the exact solution of section IV.

\section{Conclusion}

In the present work we obtained the exact solution of a probabilistic cellular automaton  related to the diagonal-to-diagonal 
transfer matrix of the asymmetric six-vertex model in a square lattice. 
The model describes the flow of ants (or particles) on a one-dimensional lattice where we have sleeping or awake ants on a ring of $L$ craters. The solution was obtained by a matrix-product ansatz that can be seen as a matrix product formulation of the coordinate Bethe ansatz. Solving numerically  the Bethe ansatz equations we calculated  the spectral gap of the model. From the  finite-size dependence of the spectral gap we verified  that the model belongs to the KPZ universality class displaying a dynamical exponent $z=\frac{3}{2}$. This 
result was also verified  from MCS. The evaluation of this exponent using MCS 
can be done  by measuring the time evolution of the standard deviation of local operators. 
 The lattice-size dependence of the decay time to the stationary state give 
us an estimator  for $z$.  Reliable results  are  
obtained by initiating the system  in a non translational invariant state. 
We can also obtain  estimates for $z$ starting with  
 translational invariant states ,provide a small local change of the 
dynamics, playing the 
rule of a "defect", is done.

\section*{Acknowledgments}

This work was supported in part by CNPq, CAPES, FAPESP and FAPERGS, Brazilian founding agencies.

\end{document}